\documentclass[a4paper,usenatbib,fleqn]{mnras}

\makeatletter

\newcommand{\Rmnum}[1]{\expandafter\@slowromancap\romannumeral #1@}
\makeatother

\usepackage{graphicx}   
\usepackage{amssymb}    
\usepackage{subcaption}
\usepackage{color,soul}
\usepackage{xcolor}
\usepackage{newtxtext,newtxmath}
\usepackage{float}
\restylefloat{table}
\usepackage{dblfloatfix}
\usepackage{booktabs}
\usepackage{multirow}
\usepackage{amsmath}    

\usepackage{color}

\def\apj{{\rm ApJ}}
\def\aj{{\rm AJ}}

\def\mnras{{\rm MNRAS}}

\def\hmpc{h^{-1}{\rm Mpc}}

\def\kms{\, {\rm km}\, {\rm s}^{-1}}

\def\lya{Ly$\alpha$ }

\def\simlt{\lower.5ex\hbox{$\; \buildrel < \over \sim \;$}}
\def\simgt{\lower.5ex\hbox{$\; \buildrel > \over \sim \;$}}

\title[Deep Forest]{
  Deep Forest: Neural Network reconstruction of intergalactic medium temperature.
}

\author[Wang et al.]{\parbox{18cm}{Runxuan Wang$^{1,2}$, 
Rupert A.C. Croft$^{1,2}$\thanks{E-mail: rcroft@cmu.edu},
and Patrick Shaw$^{1,2}$
}\vspace{0.3cm}
\\
$^{1}$ McWilliams Center for Cosmology, Dept. of Physics, 
Carnegie   Mellon  University, Pittsburgh, PA 15213, USA
\\
$^{2}$ NSF AI Planning Institute for Physics of the Future, 
Carnegie   Mellon  University, Pittsburgh, PA 15213, USA
\\
}

\begin{document}

\topmargin=-1.0cm

\maketitle


\begin{abstract}

We explore the use of Deep Learning to infer the temperature of the intergalactic medium from the transmitted flux in the high redshift \lya forest. We train Neural Networks on sets of simulated spectra from redshift $z=2-3$ outputs of cosmological hydrodynamic simulations, including 
high temperature regions added in post-processing to approximate bubbles heated by Helium-II
reionization. We evaluate how well the trained networks are able to reconstruct the temperature from the effect of Doppler broadening in the simulated input \lya\ forest absorption spectra. We find that
for spectra with high resolution (10 $\kms$ pixel) and moderate signal to noise (20-50), the neural
network is able to reconstruct the IGM temperature smoothed on scales of $\sim 6 \hmpc$ quite well.
Concentrating on discontinuities we find that high temperature regions of width 
$25 \hmpc$ and temperature $20,000$K can be fairly easily detected and characterized. We show an example where multiple
sightlines are combined to yield tomographic images of hot bubbles. Deep Learning techniques
may be useful in this way to help us understand the complex temperature structure of
the intergalactic medium around the time of Helium reionization.

\end{abstract}

\begin{keywords}
Cosmology: observations, methods: statistical, (galaxies:) quasars: absorption lines
\end{keywords}

\section{Introduction}
\label{intro}

At redshifts $z>2$ and densities relevant to the Hydrogen \lya\ forest (\citealt{rauch98}),
the temperature of the intergalactic medium (IGM) is largely governed by
photoheating from radiation backgrounds and adiabatic cooling (\citealt{hui97,miralda94}) .
Hydrogen and Helium-I reionization by radiation from the first galaxies
raised the temperature at the mean density to $\sim10^{4}$ K at redshifts
$z\sim 6-10$ (see e.g., \citealt{mquinn16} for a review) . The competition
between adiabatic cooling and photoionization heating is expected to
lead to a roughly power-law temperature-density relation (\citealt{hui97,mcquinnus16}). At
a lower redshift, $z\sim3-4$, the harder photons (with energies
$> 4$ Rydbergs) from quasars are numerous enough to reionize Helium-II
(\citealt{madau99,faucher20}). This will have lead to the last major episode of heating
of the large-scale IGM, with regions close to quasars being heated first,
causing a large scale (tens of megaparsec)
bubble-like structure of hotter regions
which then cool down (e.g, \citealt{laplante17,upton20}). The spatial structure of the IGM temperature
during this
process is likely to be relatively complex, with some regions being initially
hotter than others, then becoming cooler. Being able to infer the temperature
of the IGM at different points in space would enable us to better
understand the formation and properties of galaxies and quasars just before so-called
cosmic noon (\citealt{florez21}) and how their influence propagates to later times.
In this paper we propose a new method to do this based on Deep Learning
techniques (\citealt{lecun15,goodfellow16}) and carry out some preliminary tests using cosmological
simulations of the IGM.

The temperature of the IGM can be measured using the effect of Doppler
broadening on the forest of \lya\ absorption line seen in the spectra
of high redshift quasars and galaxies (\citealt{rauch98,savaglio02}. Traditionally, the
absorbing clouds were treated as being distinct and fitting Voigt
profiles (e.g., \citealt{vandehulst47,carswell14,webb21}) to features seen in spectra leads to two parameters,
column densities and temperatures (e.g., \citealt{hu95}). However, since the identification
of the \lya\ forest in hydrodynamic simulations (\citealt{cen94,zhang95,hernquist96}) it was realized
that in the context of cosmological models for structure formation, the
\lya\ forest is due to a fluctutating, continuous IGM (e.g., \citealt{bi93,weinberg03}). Because of
this, Hubble flow is the major source of broadening, and fitting the
width of features with a thermally broadened profile of a discrete object
will not yield the temperature directly. Many other techniques have
therefore been developed to constrain the IGM tempetature from the \lya\ forest.
One is the relatively straightforward use of Voigt profile fit
parameters to try to find the envelope which corresponds to a minimum
IGM temperature (e.g, \citealt{garzilli20}). Others include wavelet transforms (\citealt{meiksin00,garzilli12}), or measuring
statistical properties of the \lya\ forest flux, such as the power spectrum
(\citealt{croft98,lai06}), which respond to the small scale smoothing that results from
increased IGM temperatures. More recent ideas include the reconstruction
techniques of \cite{muller21} that use 3D tomographic data. In the current
work we will use a Neural Network (NN, see below) trained on simulated
\lya\ spectra to infer the temperature from the smoothness of
absorption features. We will be looking in particular for
discontinuous $T$ jumps which could result from discrete sources of
heating such as quasars reionizing HeII. We note that the temperature
of gas will also influence the pressure smoothing scale (\citealt{gnedin98}), which physically
makes gaseous structures smoother than their underlying dark matter
counterparts \citep{peeples10}. In our current preliminary work we will not
model this effect (see also the approach of \citealt{mcquinn11}, who used a similar
approximation), so that the
smoothing of absorption
features will be entirely attributed to Doppler broadening.

Deep Learning (DL, see e.g., the review by \citep{goodfellow16}, a branch of Machine Learning (ML) which involves
the use of artificial NNs has become
extremely popular in astronomy and many other fields of science.
An overview of ML in astronomy is \cite{baron19}.
The NNs
consist of neurons arranged in layers which communicate via weighted
connections, those weights being adjusted as the NN is trained. Their use
in astronomy started mainly as classfiers of events,
images (e.g., \citealt{cheng20}), and also spectra (\citealt{muthukrishna19}) but they are
increasingly being used in analysis (e.g.,
of gravitational wave time series data, \citealt{george2018}, spectral line fitting, \citealt{lee21}), and to enhance and speed
up numerical simulations (e.g., \citealt{ramanah20,li20}). Cosmological N-body
simulations (\citealt{vogelsberger20}) have been used as training sets for NNs, for example
in the generation of images of structure formation (\citealt{caldeira2019}), or
to incorporate galaxy properties into gravity-only N-body model (\citealt{lovell21}).
DL is also useful as a tool for reconstruction and analysis (see e.g., \citealt{laplante19,hassan19} for
applications to maps of reionization). In \cite{huang21}
(hereafter H21) it was shown how the underlying neutral hydrogen density
could be reconstructed from \lya\ forest spectra, using NN trained using
cosmological simulations to learn the relationship between the two. We will
be using the same simulations here, and a similar NN architecture to
H21, but will instead focus on the temperature of the IGM.

Our plan for the paper is as follows. In Section \ref{simandpreprocess}, the parameters and basic statistics in the hydrodynamic simulation used are introduced, along with preprocessing of the training data. In Section \ref{method} we present the architecture of the Neural Network model, details of training including the selection of loss function, and  hyperparameter tuning. Results of the training data are first numerically shown in tables and then visualised in 2D plots in Section \ref{results}, with the analysis of the model robustness to noise and generalisation across
redshifts. In the last section, Section \ref{DiscusssionandConclusions}, we present our conclusions, discuss the limitations of the study, and  potential improvements in architecture, training, and
other future work.

%


\section{Hydrodynamic Simulation Outputs and Data Preparation}
\label{simandpreprocess}
\subsection{The Hydrodynamic Simulation}
\label{sim}

A cosmological hydrodynamic simulation was used to create a set of simulated \lya\ forest spectra for training 
the NN and for tests. The simulation was run using $2 \times 4096^2 = 137$ billion particles with the
smoothed particle hydrodnamics (SPH) code, {\small P--GADGET}  (see \citealt{springel05,dimatteo12}) in a $(400 \hmpc)^3$ cubic volume.  The 
underlying cosmological model was $\Lambda$CDM with parameters $h = 0.702,\, \Omega_{\Lambda} = 0.725,\, \Omega_m = 0.275, \, \Omega_b = 0.046, \, n_s = 0.968\; $and$ \; \sigma_8 = 0.82.$. More details on the
simulation are given in \cite{cisewski14}, \cite{croft18} and H21, where it was also used to make mock \lya\ forest spectra for analysis. 
Simulation outputs were available at three redshifts, $z = 3.0, 2.5, 2.0$; the first two were used to train the NN.

The spectra themselves were constructed using the standard techniques (e.g., \citealt{hernquist96}) relevant for SPH simulations. Briefly, the neutral hydrogen density was assigned to
spectral pixels using the SPH kernels of particles that intersect them. The same is true of 
other physical quantities, which are weighted by the neutral hydrogen density (proportional to
the optical depth for absorption).
 Along each line of sight, the simulation spectra data include the following physical quantities which
 we use in our modeling: optical depth
 of neutral hydrogen absorption in real space $\tau_\textrm{real}$, peculiar velocity component along
 the given sightline $v$, temperature $T$, and normalised baryon density in units of the mean density $\rho$. 
 The conversion of $\tau_\textrm{real}$ into the observable quantity, the transmitted flux in a spectrum, $F$,
 involves convolution with the peculiar velocity together with thermal broadening. Because we manipulate
 the spectral data by inserting hot bubbles of different types as part of our training and
 testing, we describe this process later.
 
 Each spectrum has a length equal to the box length and is discretized into 4096 pixels with size $dz \approx 10 \kms$. More exact quantitative information is shown below in table \ref{siminfo} which gives the sight line length and pixel sizes for the data at different redshifts.

\begin{table}
    \centering
    \caption{Information on the simulated spectra used}
    \label{siminfo}
    \begin{tabular}{llll}
        \hline
        Name & $z=3.0$ & $z=2.5$ & $z=2.0$ \\
        \hline
        Number of Sightlines & 65536  & 65536 & 65536  \\
        Resolution (Pixel Number) & 4096 & 4096 & 4096 \\
        Box Side Length ($10^{4}\kms$) & 4.28 & 4.04 & 3.80 \\
        Pixel Size ($\kms$) & 10.45 & 9.87 & 9.29 \\
        \hline
    \end{tabular}
\end{table}

\begin{figure}
    \includegraphics[width=\linewidth]{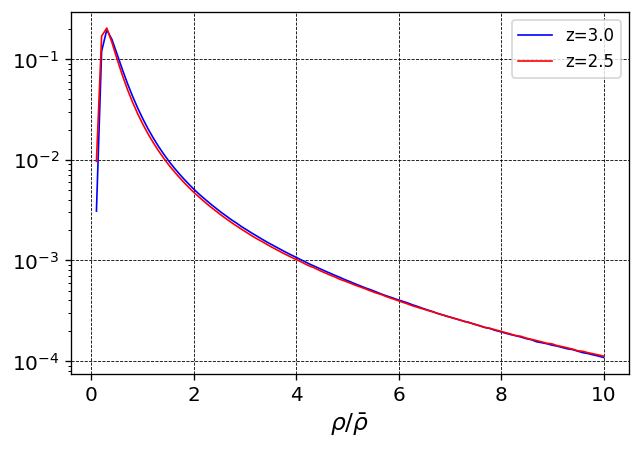}
    \includegraphics[width=\linewidth]{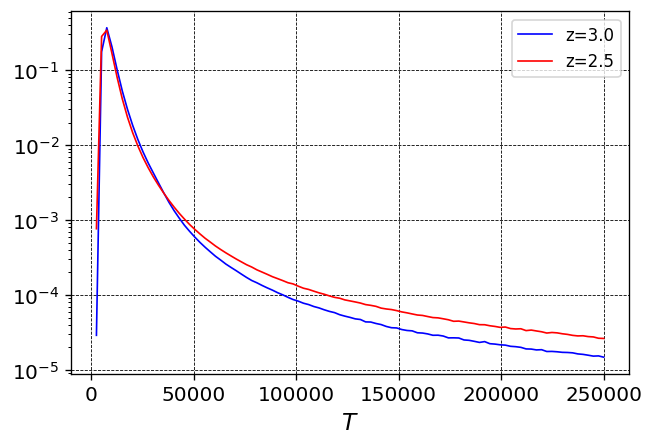}
    \caption{The top histograms shows the distribution of density in simulated \lya\ forest for redshift $z=3.0$ and $z=2.5$; the bottom histogram shows the distribution of temperature.}
    \label{stats}
\end{figure}

The two plots in Figure \ref{stats} show the probability distributions of the density and temperature pixel values
in the two datasets used to train the NN. We can see that $\rho$ and $T$ have similarly skewed distributions,
as would be expected given relation between temperature and density expected in the
IGM relevant for the high $z$ \lya\ forest (e.g., \cite{hui97}).  
Over $90\%$ of the pixels have temperature less than approximately $25,000$ K and density less than $1.5$.

\begin{figure*}
    \includegraphics[width=\linewidth,height=0.21\linewidth]{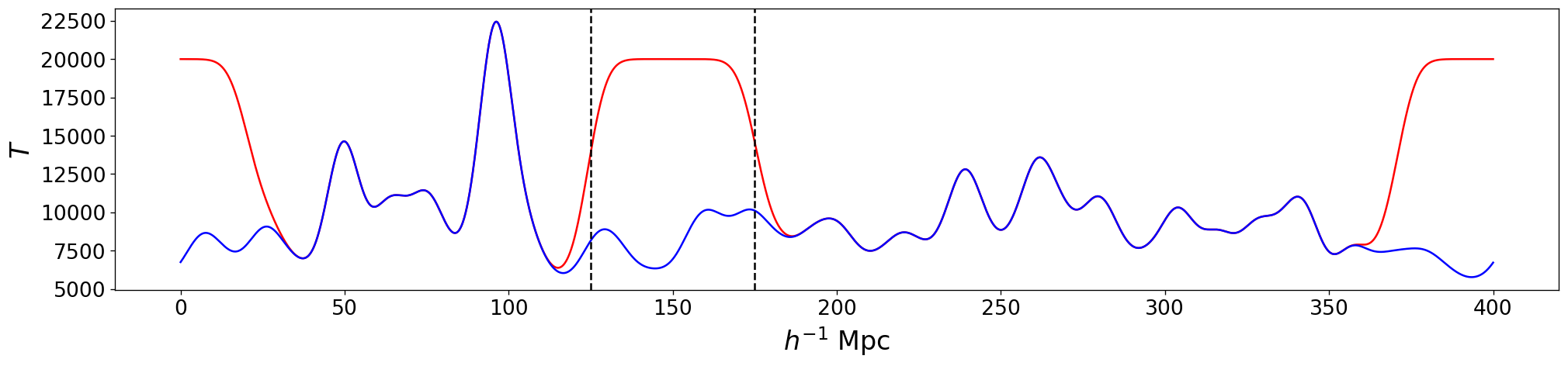}
    \includegraphics[width=\linewidth,height=0.21\linewidth]{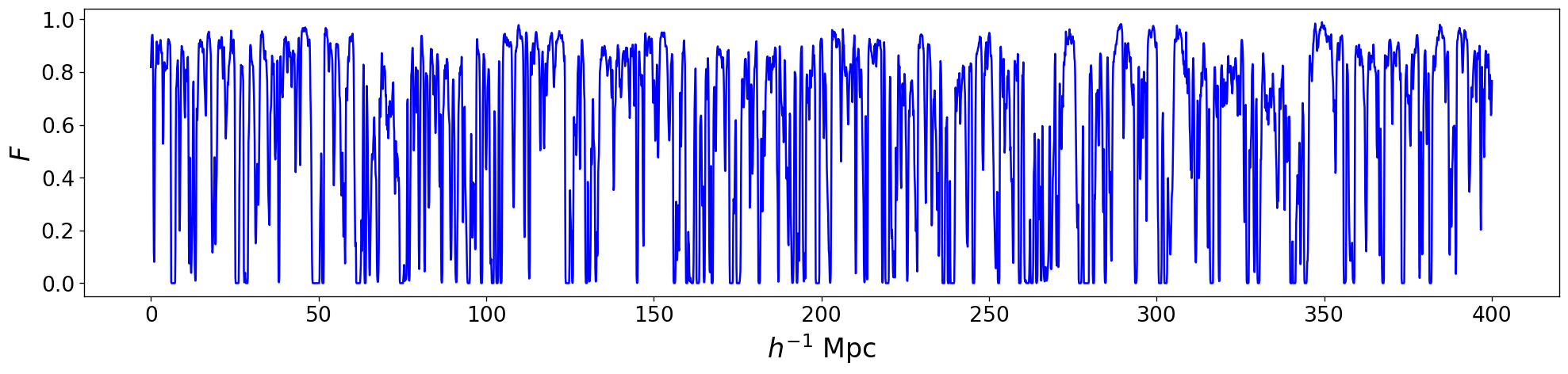}
    \includegraphics[width=\linewidth,height=0.21\linewidth]{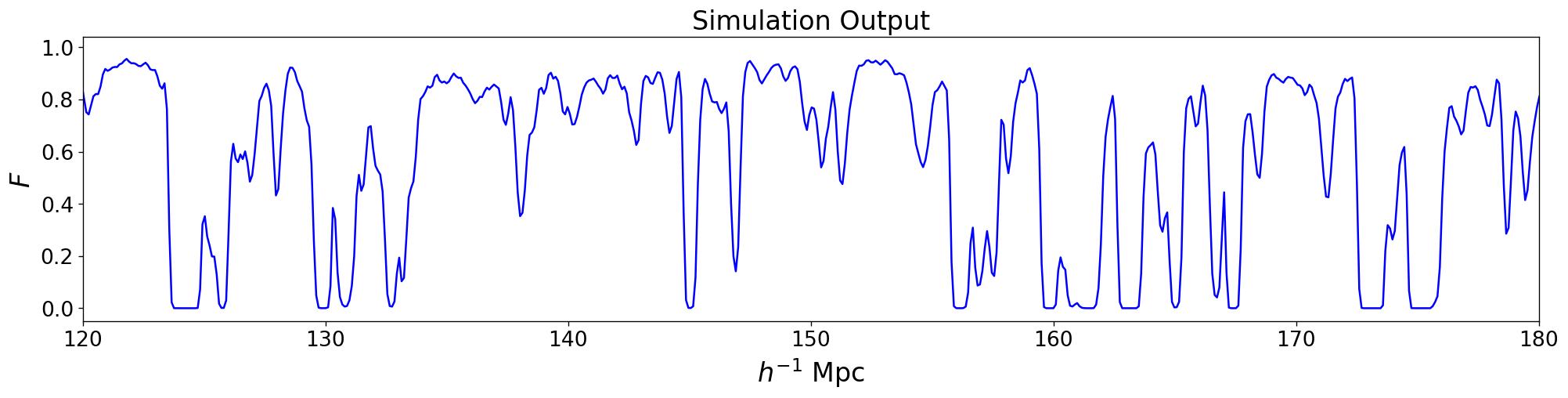}
    \includegraphics[width=\linewidth,height=0.21\linewidth]{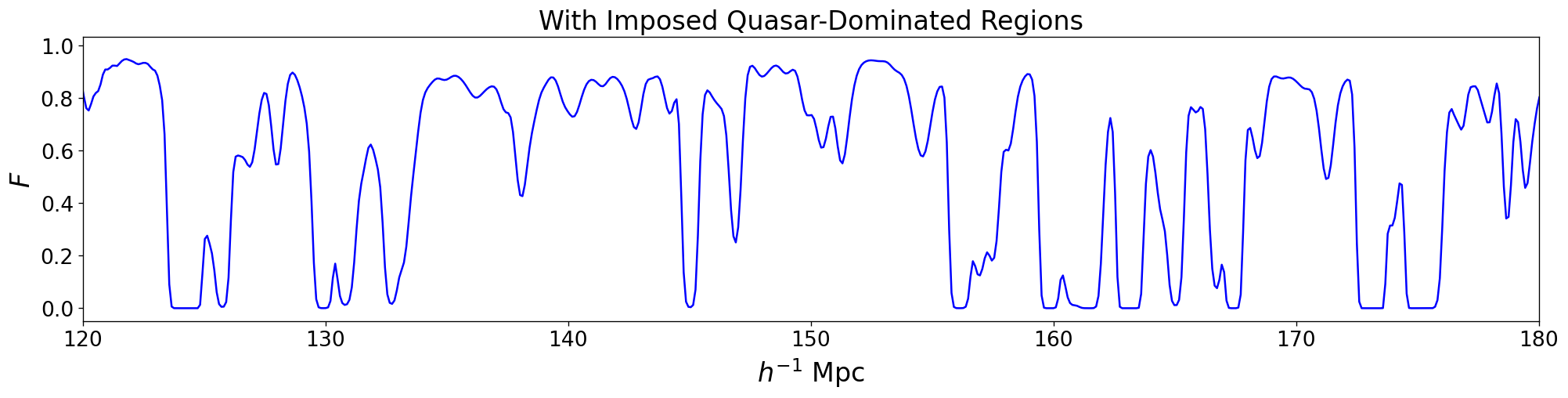}
    \caption{The spectrum of the observed flux $F$ and temperature $T$ along the same sightline.  The information on the number of pixels and resolution is given in Table \ref{siminfo}. The temperature (top panel) has been smoothed using a Gaussian filter of kernel size of $4.7 \hmpc$; the transmitted flux spectrum (not smoothed) is shown in the second panel. The red bulges are mock quasar-dominated regions (mentioned in section \ref{preprocess}) imposed in the training sets. A comparison between the transmitted flux with (bottom panel) and without (third panel from top) the imposed high temperature region is shown in the bottom two panels in a zoomed interval between $120 \hmpc$ and $180 \hmpc$. }
    \label{exFT}
\end{figure*}

The transmitted flux $F$ and temperature $T$ along an example sightline at redshift $z=3.0$  are shown in Figure \ref{exFT}.
In order to make the large scale structure of the $T$ field more apparent, we first smooth the temperature distribution along the sightline with a Gaussian filter of kernel size  $4.7 \hmpc$, which is approximately $1.2\%$ of the box length.
The red line in the temperature panel corresponds to
the added mock quasar-dominated regions in the processed spectra (mentioned in section \ref{preprocess}) where the gas has been heated above the level computed by the
initial simulation. In the bottom two panels we show
 the flux at full resolution zoomed in to make smaller structures easier to see.
A comparison between the transmitted flux with (bottom panel) and without (third panel from top) the imposed high temperature regions is shown in the bottom two panels in a zoomed interval. We can clearly see that the more intense Doppler broadening in the quasar-dominated region (which covers the entire zoomed-in bottom panel) has the effect of smoothing the flux.

\subsection{Training Data Preparation}
\label{preprocess}
The simulation output at each redshift exhibits a strong correlation between $\rho$ and $T$ in each pixel (\citealt{hui97}). A 2D histogram of each pixel's density and temperature is shown below in Figure \ref{rhoT}, color-coded by the density of points.

\begin{figure}
    \centering
    \includegraphics[width=0.8\linewidth]{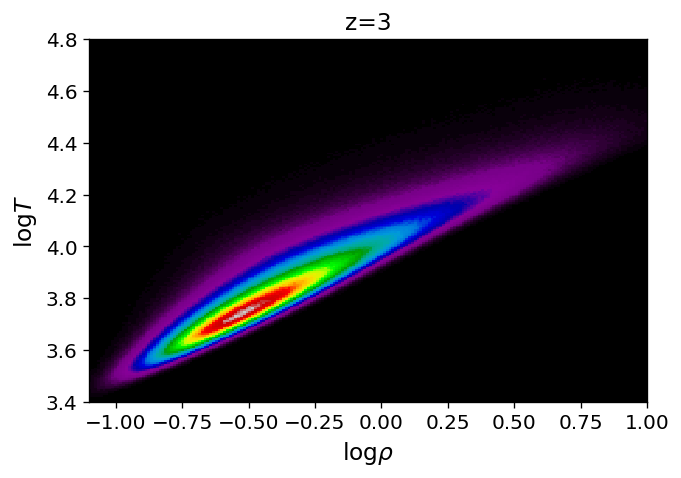}
    \includegraphics[width=0.8\linewidth]{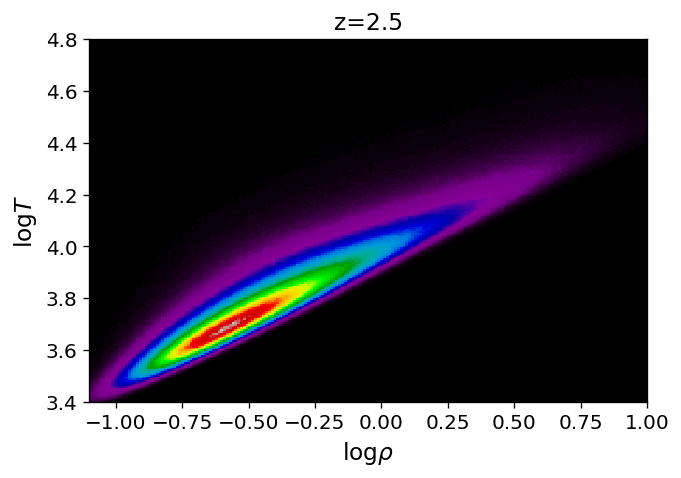}
    \caption{2D Histogram of Density and Temperature in the simulation output for $z=3$ (top) and $z=2.5$ (bottom). }
    \label{rhoT}
\end{figure}

The relationship can be fit to a power law: for $z=3.0$, we find $T = 11600\rho ^ {-0.54}$ K. Temperatures from real observations could be more extreme and spatially inhomogeneous,  not following the same  uniform power-law relation everywhere (see e.g., \citealt{upton20}). Because of this, our approach with a NN is to avoid 
inferring the temperature from the observed flux $F$ (and hence the density), but rather to use information
from the smoothness (thermal broadening) alone. In our training we therefore decouple the temperature values from
the density, assigning  different temperatures from those computed by the hydrodynamic simulation and then
ensuring that the thermal broadening is consistent with these new temperatures.

To achieve this, we tried two methods to randomise the $\rho-T$ relation. 
One method is to discard the simulation temperature and instead for every sightline draw randomly and independently from a probability function of temperature. This is mentioned in more detail in Section \ref{discussionTraining}. 
The other method involves imposing randomly placed quasar-dominated regions with high, constant $T$ on the original spectrum at random locations. For each sightline, we add two quasar-dominated patches in which every pixel has the same temperature at random locations. 
This is a simple model of the physical effect of hard spectrum of ionising radiation from a quasar that can ionise Helium (e.g., \citealt{sokasian02}). The two parameters are the constant temperature and the width of the patch; the patch temperature is set to be $20,000$ K, $30,000$ K, or $50,000$ K, and the patch width is set to be $25 \hmpc$, $50 \hmpc$, or $67 \hmpc$, corresponding to $1/16$, $1/8$, or $1/6$ of the sightline length.

We also include the effects of varying signal to noise in the spectra. We add three different levels of simulated observational noise to the $F$ values in pixels by drawing from a Normal distribution with $\sigma$ equal to the signal to noise ratio $S/N=20$ and $S/N=50$, so that $F_\textrm{noised} = F + \textrm{Normal}(0,S/N)$.

To train the Neural Network to reconstruct temperature from \lya spectra, the input of the training set is the simulated observed flux $F$, which we recalculate (remember that we are imposing
a new set of temperatures) in the following way. First, the optical depth in redshift space, $\tau_\textrm{red}$, is computed by accounting for the Doppler broadening effect due to peculiar velocity and thermal motion of the gas. This convolution with the velocity field and thermal broadening is carried out as follows: $\tau_\textrm{red}$ is obtained by displacing $\tau_\textrm{real}$ at each pixel by an amount equal to the peculiar velocity in that pixel  while at the same time the delta function value of $\tau_\textrm{real}$ in the pixel is replaced by a Gaussian distribution with standard deviation proportional to the square root of the temperature. This last value is  $\sigma=0.12849 \sqrt{T} \kms$, 
appropriate for hydrogen. The resulting mock observed flux in each pixel is then $F = e^{- \tau_\textrm{red}}$.

Due to limited GPU memory, we use only 30,000 out of the available 65,536 sightlines for each run of training. The training set has 20,000 sightlines, the validation set has 5000, and the test set has 5000. Since the three-dimensional structure of sightline within the box will lead to correlations between sightlines that are spatially near, the training, validation and test sets are drawn from 
separate intact cuboids in the simulation box.

\section{Reconstruction Method}
\label{method}

\subsection{Neural Network Architecture}
\label{NN}
\begin{figure*}
    \includegraphics[width=\linewidth]{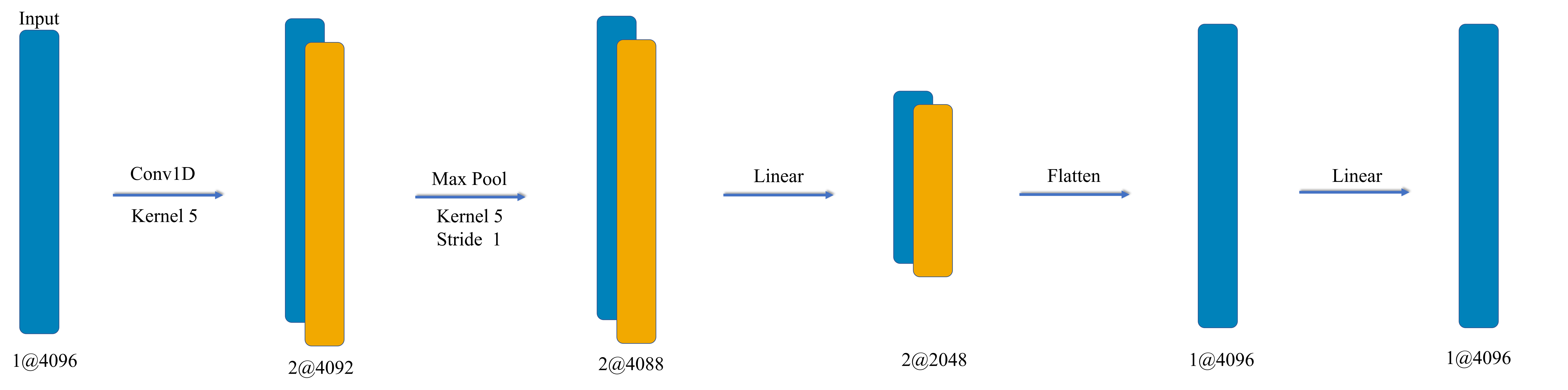}
    \caption{The architecture of the Neural Network model. The input: individual transmitted flux spectrum (size 1$\times$4096). Layer 1: one-dimensional convolution layer of size 4096 pixels (kernel size 5 pixels); outputs tensor (size 2$\times$4092). Layer 2: Max Pool layer (kernel size 5 pixels stride 1); outputs tensor (size 2$\times$4088). \\Layer 3: linear layer; outputs tensor (2,2048), which is then flattened into size 1$\times$4096. Layer 4: linear layer; outputs temperature spectrum of 1*4096. Activation function coupled with layer 1 and layer 3: Scaled Exponential Linear Unit (SELU), where $\textrm{SELU}(x) = \lambda (\textrm{max}(0,x)+\textrm{min}(0, \alpha (e^x - 1)))$.}
    \label{archi}
\end{figure*}
The NN approach we use will take as an input the transmitted \lya\ forest flux $F$ in pixels
for a sightline, and output the
temperature values in the same pixels. To achieve this, the NN will be trained on mock spectra with
a range of different temperatures so that it learns to associate features of the flux $F$ that are 
caused by thermal broadening with the underlying gas temperature.

The architecture of the NN model we developed is shown in Figure \ref{archi}. The overall architecture is inspired by the work of H21 and implemented using Pytorch\footnote{\tt pytorch.org}. The input is the observed (with or without noise added) transmitted flux spectrum (size 1 $\times$4096) for an individual line of sight. The first layer takes in the flux spectrum through a one-dimensional convolution layer of size 4096 pixels, and outputs two channels (size 2$\times$4092). The second layer is the Max Pool layer with kernel size of 5 pixels and default stride (1), which extracts the maximum value of consecutive patches of 5 pixels, and outputs a tensor of size 2$\times$4088. Then, a linear layer reduces the number of neurons by about a half. The resulting tensor is then flattened into a single channel, before entering the final layer which maps the tensor into the output temperature spectrum of equal size (1$\times$4096). Both after the 1-D convolution layer and the first linear layer, the activation function used is Scaled Exponential Linear Unit (SELU), defined by $\textrm{SELU}(x) = \lambda (\textrm{max}(0,x)+\textrm{min}(0, \alpha (e^x - 1)))$. $\lambda \approx 1.051$ and $\alpha \approx 1.673$ are taken to be the default values.

The architecture is chosen mostly by intuition and experimentation. To account for the correlations within the nearby pixels in a sightline, the convolution layer is included. This model only has two linear layers, one less than similar models with similar architectures, which we have also trained and tested for reconstruction of other physical quantities (e.g., density and optical depth) which contain less extreme values. We find that, when reconstructing temperature, the addition of any linear layer introduces too many parameters causing the model to overfit, even with dropouts.   Therefore, the use of a linear layer might be a reasonable maximum for temperature reconstruction.

To avoid overfitting, we attempt to apply dropout (\citealt{srivastava14}) at various places between the linear layers.
Dropout is a regularization method that approximates training a large number of NN with different architectures in parallel.
During training, some number of layer outputs are dropped out, or randomly ignored, 
but test runs show that for this model dropouts below $30\%$ of the neurons do not affect accuracy by much, and dropouts above $30\%$ of the neurons  only lead to failure to converge.

\subsection{Training}
\label{training}

The loss function that gives the best results is the mean absolute error (or L1 loss) of the prediction and the actual temperature spectrum, simply defined as
\begin{align}
L = \frac{1}{N} \sum_{i=0}^{N} \sum_{n=0}^{4096} \left|T_{p,in} - T_{a,in}\right| ,
\label{loss}
\end{align}
where N is the number of lines of sight in a batch (batch size), the $T_{p,i}$ and $T_{a,i}$ are the predicted and actual temperature of the $i^{\textrm{th}}$ sightline, respectively.

Two of the alternative loss functions we have attempted to use are the mean square error (MSE) and root mean square log error (RMSLE), defined below:
\begin{equation} \label{MSE}
L_{\textrm{MSE}} = \frac{1}{N} \sum_{i=0}^{N} \sum_{n=0}^{4096} \left(T_{p,in}-T_{a,in}\right)^2,
\end{equation} 
\begin{equation}
L_{\textrm{RMSLE}} = \frac{1}{N} \sum_{i=0}^{N} \sum_{n=0}^{4096} \left(\log({T_p}_{in}+1) - \log({T_a}_{in}+1) \right)^2.
\end{equation}
MSE and RMSLE behave well for physical quantities with smaller values such as optical depth and density, which typically lie within the range $(0,15)$. MAE is more robust to extreme values of high temperature ($>100,000$ K); it prevents the model from drifting in the parameter space too quickly upon encountering  several high temperature pixels.

The learning rate is set to be $10^{-3}$. Our test runs show that learning rates less than $10^{-3}$ mostly fail to converge. Unlike depth and density reconstruction, for which the learning rate is usually $10^{-4}$, the scale for the temperature values in sightlines is about three orders of magnitude larger, and nearly every sightline has a peak that extends above two times the mean, so that the lower limit of the learning rate for our model is about $~10^{-4}$. The upper limit is approximately $5\times10^{-2}$, given that for our test runs learning rates above this rarely lead to convergence. The Pytorch StepLR scheduler that decreases the learning rate exponentially with the number of epochs was tried, but its effect is negligible.

The batch size is chosen to be 3000 to 4000. The best performing batch size varies as the width and temperature of the mock quasar-dominated regions change. Larger batch sizes are preferred for small widths and low temperatures and vice versa. The optimiser used is the Adam optimiser with a weight decay of $0.0005$; the choice of the weight decay came from experimentation. 

As described in Section \ref{preprocess}, we train the model, with the above architecture and hyper-parameters, for various temperature distributions by artificially imposing constant high temperature patches, a toy model for a quasar-heated region. The input is the recalculated transmitted flux with three possible noise levels. The number of epochs in training vary between $10,000$ to $20,000$. For different variations of the patches, if we initialise the parameters of the model using an already trained model for a patch of different temperature and width, we find that about $4000$ epochs are necessary before convergence.


\section{Results}
\label{results}
\subsection{Reconstructed  Energy Ratio}
\label{energyRatio}

The optical depth $\tau_{\textrm{real}}$ in real space is strongly correlated with $\tau_{\textrm{red}}$ in redshift space, which is the quantity used to directly compute the observed flux: $F = e^{- \tau_{\textrm{red}}}$. The density $\rho$ and $\tau_{\textrm{real}}$ are also correlated with an approximate power law relationship holding between them, 
and we have already seen (Figure \ref{rhoT}) how $\rho$ and $T$ are related.
We need to make sure that the NN does not learn these relationships directly, because if
it did, then when applied to an unknown data sample the easiest path to estimating the
$T$ values in pixels from $F$ would be to invert the power law relationships mentioned.
We need to decouple the temperature from  the density in the training set.
Thus, as discussed in section \ref{preprocess}, we randomly assign the location of the
high temperature patches in training independent of the density.
Because of this the NN will learn to reconstruct $T$ from the Doppler/thermal broadening effect alone, without assuming a uniform $\rho-T$ relation.

We restrict our quantitative evaluation of the accuracy of the NN reconstruction of temperature to  only  the mock quasar-dominated regions. We define a statistical measure of the agreement between the temperature predicted by the NN and the true $T$ values for an individual sightline to be 
\begin{equation}
    s = \frac{E}{E_0} = \frac{\int_{x_1}^{x_2} T_p(z) dz}{(x_2-x_1)T_c},
\end{equation}
where $T_c$ and $T_p$ are the constant temperature of the patch and the predicted $T$ values
in pixels at position $z$ respectively. Here $w$ is the patch width (we
try three possible values, $25,50$ and $67 \hmpc$). $x_1$ and $x_2$ are the left and right boundaries of the reconstructed patch; starting from the centre $x_c$ of the region and extending in both directions, $x_1<x_c$ is the point where the predicted temperature $T$ becomes less than the mean of the reconstructed spectrum for the first time, and similarly for $x_2>x_c$. The patch width of the prediction is then $w_p = w_2 - w_1$. $E$ and $E_0$ are the actual and the predicted integral of temperature along the distance from the observer (equivalent to an integral over wavelength). $E$ and $E_0$ therefore represent the internal thermal energy of the quasar-dominated region (and are exactly proportional to
it in the case of a uniform gas density across the patch). Their ratio is a measure of how well the NN has recovered the temperature in a heated region.

\begin{figure}
    \includegraphics[width=\linewidth,height=0.4\linewidth]{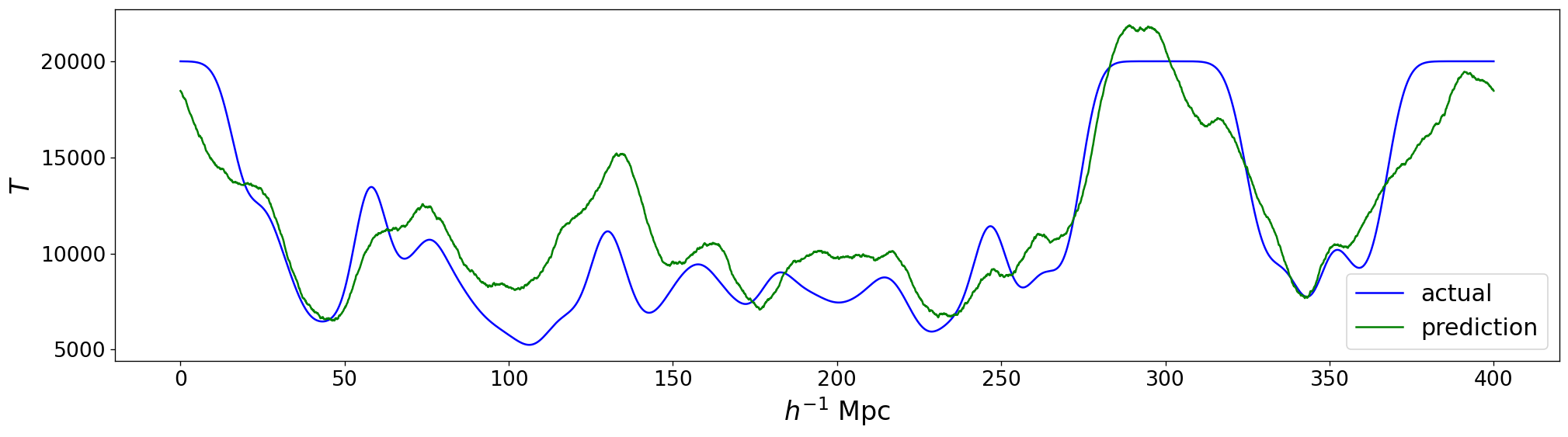}
    \includegraphics[width=\linewidth,height=0.4\linewidth]{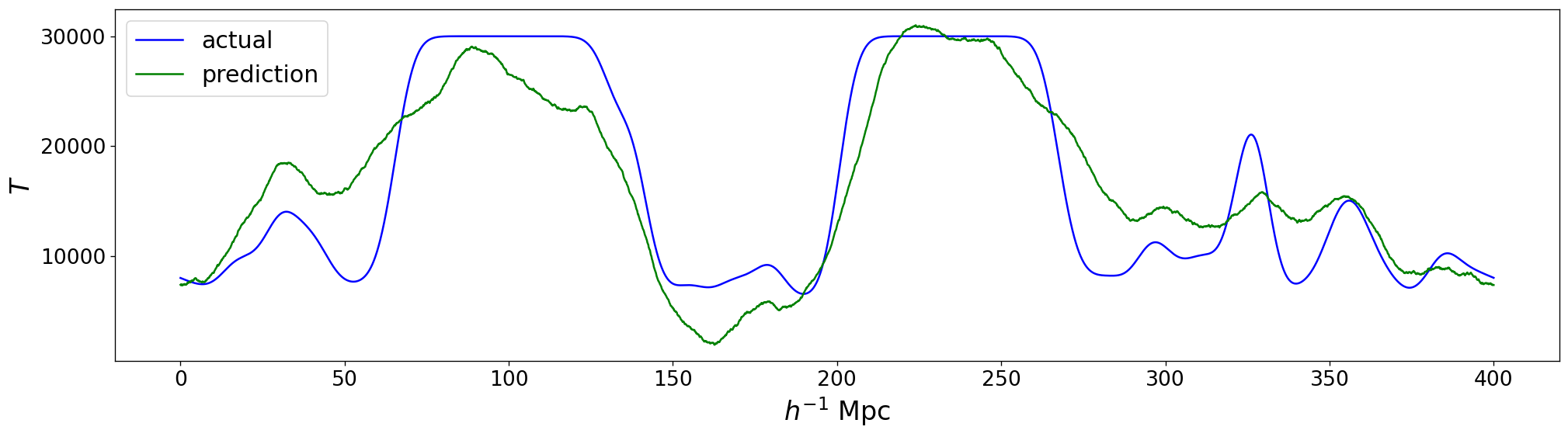}
    \caption{Example sightline with actual and reconstructed $T$ values  (smoothed) for $T_c=20,000$ K and $w=50\hmpc$ (top) and $T_c=30,000$ K and $w=67 \hmpc$ (bottom)}
    \label{exampleConstQuasar}
\end{figure}

Figure \ref{exampleConstQuasar} shows two examples of the comparison between the predicted and actual temperatures along a sightline. The top example has $w=50 \hmpc$ and $T_c = 20,000K$, and the bottom case  $w=67 \hmpc$ and $T_c = 30,000K$. By visual inspection of these results we infer that the Neural Network is able to recover the position $x_c$ of the high temperature region quite well. We also see that for the pixels far from the patch region the prediction has a larger fractional error. Periodic boundary conditions are applied so that there is no bias in the result related to the relative position 
of the patch in the spectrum. 

We test the NN $T$ reconstruction on 5000 examples each of several different combinations of patch temperature, width, snapshot redshift and noise level.
For each we compute $s = E/E_0 $ values, and
record the mean $s$ value (which serves to identify overall biases in the reconstructed temperature)
and the standard deviation about that mean (to quantify the expected scatter in reconstruction from patch to patch)
Table \ref{tab:result1} and table \ref{tab:result2}  show the $s = E/E_0 $ values for the $z=3.0$ and $z=2.0$ simulations, respectively. The  rows labelled with $S/N$ show the noise level. Within each noise level, labels on the rows are the patch temperatures $T_c$ in units of $10^3$ K. The column labels represent the patch width $w$ in $\hmpc$.

\begin{table}[htbp]
  \centering
  \caption{$E/E_0$ for $z=3.0$}
    \begin{tabular}{|c|c|c|c|c|}
    \toprule
    $z=3.0$    &   $T_c (10^3 \textrm{ K})$  & 25$\hmpc$ & 50$\hmpc$ & 66.67$\hmpc$ \\
    \midrule
    \multirow{3}[6]{*}{S/N=inf} & 20 & 0.82±0.37 & 1.17±0.21 & 1.09±0.15 \\
\cmidrule{2-5}          & 30 & 0.96±0.27 & 0.92±0.11 & 1.06±0.10 \\
\cmidrule{2-5}          & 50 & 1.2±0.21 & 1.09±0.09 & 1.02±0.07 \\
    \midrule
    \multirow{3}[6]{*}{S/N=50} & 20 & 0.76±0.36 & 1.12±0.23 & 1.07±0.17 \\
\cmidrule{2-5}          & 30 & 0.88±0.28 & 1.15±0.14 & 1.06±0.10 \\
\cmidrule{2-5}          & 50 & 0.86±0.18 & 1.09±0.10 & 1.02±0.07 \\
    \midrule
    \multirow{3}[6]{*}{S/N=20} & 20 & 0.69±0.35 & 1.02±0.26 & 1.03±0.20 \\
\cmidrule{2-5}          & 30 & 0.76±0.28 & 1.13±0.17 & 1.06±0.12 \\
\cmidrule{2-5}          & 50 & 0.74±0.18 & 1.08±0.12 & 1.02±0.08 \\
    \bottomrule
    \end{tabular}%
  \label{tab:result1}%
\end{table}%

\begin{table}[htbp]
  \centering
  \caption{$E/E_0$ for $z=2.5$}
    \begin{tabular}{|c|c|c|c|c|}
    \toprule
    z=2.5 &   $T_c (10^3 \textrm{ K})$    & 25$\hmpc$    & 50$\hmpc$   & 66.67$\hmpc$ \\
    \midrule
    \multirow{3}[6]{*}{S/N=inf} & 20    & 0.61±0.33 & 1.01±0.25 & 1.05±0.19 \\
\cmidrule{2-5}          & 30    & 0.83±0.28 & 1.15±0.15 & 0.90±0.34 \\
\cmidrule{2-5}          & 50    & 0.81±0.19 & 1.09±0.10 & 1.02±0.08 \\
    \midrule
    \multirow{3}[6]{*}{S/N=50} & 20    & 0.55±0.32 & 0.87±0.26 & 0.98±0.22 \\
\cmidrule{2-5}          & 30    & 0.71±0.29 & 1.27±0.17 & 0.89±0.34 \\
\cmidrule{2-5}          & 50    & 0.71±0.19 & 1.09±0.12 & 1.02±0.09 \\
    \midrule
    \multirow{3}[6]{*}{S/N=20} & 20    & 0.51±0.31 & 0.67±0.24 & 1.18±0.27 \\
\cmidrule{2-5}          & 30    & 0.59±0.27 & 1.38±0.22 & 0.84±0.35 \\
\cmidrule{2-5}          & 50  & 0.59±0.19 & 1.07±0.15 & 1.02±0.11 \\
    \bottomrule
    \end{tabular}%
  \label{tab:result2}%
\end{table}%

\begin{figure*}
    \includegraphics[width=1.0\linewidth]{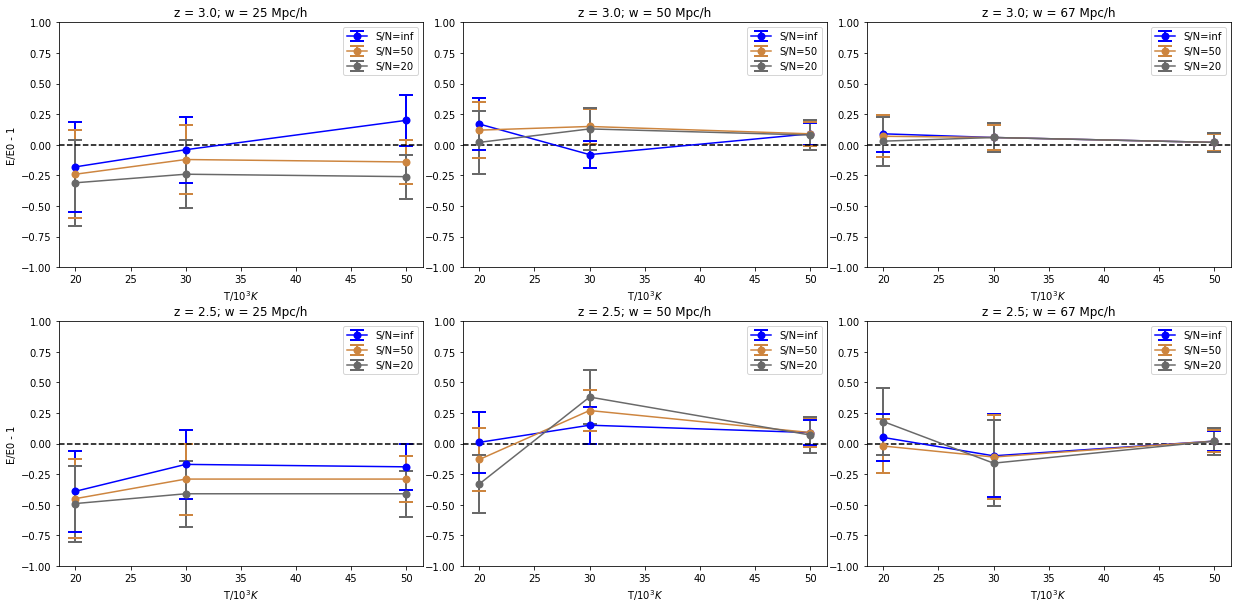}
    \par\bigskip
    \includegraphics[width=1.0\linewidth]{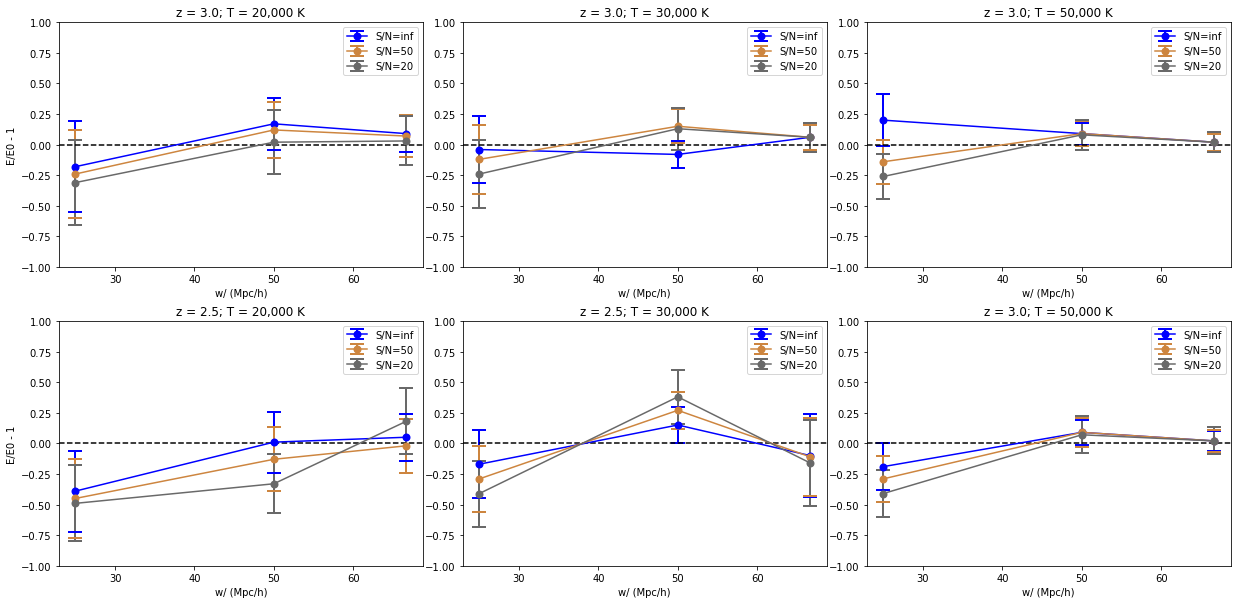}
    \caption{Accuracy of the NN reconstruction of total integrated internal energy in heated
    IGM patches. Plots are shown of $s - 1 = E/E_0 - 1$ for various combinations of patch temperature $T_c$ (the independent variable in the first two rows) and patch width $w$ (the last two rows). Three noise levels of $S/N=+\infty$ (no noise), $50$,and $20$ are included in each panel.}
    \label{lineplot}
\end{figure*}

Figure \ref{lineplot} visualizes how the accuracy (here plotting $s-1=E/E_0 -1$) of the NN varies as the patch temperature $T$ and patch width $w$ change. The independent variable for the top two rows of plots is $T$, and for the bottom two $w$; redshift and $w$ are labelled for each panel. For each plot, three differently colored lines represent the three noise levels added to the observed flux. A weaker smoothing effect due to relatively low patch temperatures and/or narrow regions are expected to be harder to detect by NN, so that the expectation is that the reconstructed energy will be closer to the actual value 
(i.e. $s\sim 0$) for larger $T_c$ and $w$ values. We also expect that the effect of noise decreases for larger $T_c$ and $w$. The following observations agree with these expectations:
\begin{enumerate}
\item{The best performing case we have tested is $T=50,000\textrm{ K}, w=67\hmpc$. For both redshifts the thermal energy can be reconstructed on average within $2\%$ with standard deviation $\sigma < 11\%$ given the noise level $S/N>20$. The condition where the model performs the worst is $T=20,000\textrm{ K}, w=25 \hmpc$, the mean reconstructed energy is on average less than $82\%$ of the actual value with a large scatter of $30\%$.}
\item{The accuracy increases with decreasing noise in most cases. The line does converge to $0$ for higher temperatures and larger width $w$. Generally, the difference in accuracy caused by noise decreases with increasing values for $T$ and $w$, the three lines spread less for every panel except for $z=3.0, w=25 \hmpc$ (the top left).} 
\item{The error, which is defined as the standard deviation of the distribution of $E/E_0$ for the 5000 sightlines in the test set, decreases with higher $T_c$ and $w$.}
\end{enumerate}

 We therefore conclude that, for mock quasar-heated regions with $w\geq 25\hmpc$ and $T\geq 20,000\textrm{ K}$, the Neural Network is able to reconstruct the position and the approximate width (shape) of those regions through the Doppler broadening effect alone from the \lya forest spectrum. The accuracy of the reconstructions are quantified by a statistical measure related to the internal thermal energy of these regions; the scores are listed in table \ref{tab:result1} and \ref{tab:result2}. 

\subsection{Test of generalizability}
\begin{table}[htbp]
  \centering
  \caption{$E/E_0$ for $z=3.0$ using NNs trained on $z=2.5$ data.}
    \begin{tabular}{|c|c|c|c|c|}
    \toprule
    $z=3.0$    &   $T_c (10^3 \textrm{ K})$  & 25$\hmpc$ & 50$\hmpc$ & 66.67$\hmpc$ \\
    \midrule
    \multirow{3}[6]{*}{S/N=inf} & 20 & 0.94±0.47 & 0.88±0.21 & 0.86±0.16 \\
    \cmidrule{2-5}              & 30 & 0.75±0.23 & 0.97±0.20 & 0.88±0.10 \\
    \cmidrule{2-5}              & 50 & 1.09±0.21 & 1.33±0.31 & 0.88±0.06 \\
    \bottomrule
    \end{tabular}%
  \label{tab:result3}%
\end{table}%

The reconstruction of temperature structure by the NN will depend on the training set. We have so far only tested the technique on spectra drawn from the same simulation (redshift,
cosmological parameters) as the training data. A general issue with NN is how well their
outputs generalize when applied to datasets that they have not seen in training. In future work it will be useful
to investigate this by training with different cosmologies or implementations of cosmological hydrodynamics (see e.g., \citealt{villa21} for efforts along these lines). In the meantime we carry out a limited version of such tests by attempting to reconstruct temperature and find
heated regions using a NN trained using a different redshift output.

 We apply the NN model trained on $z=2.5$ data with different patch widths and patch temperatures, and test them on the same situation but for $z=3.0$ spectra. No noise is added in this case. The same measure $E/E_0$ is used, and the results are shown in table \ref{tab:result3}. We might expect that overall the reconstructed energy ratio should be
 relatively different from 1 compared to training on the correct redshift. We see that large patch temperatures and a wide region are reconstructed to higher accuracy, as before.
Overall though, comparing to Table 2 we see that the errors are quite similar to the case trained  on $z=3.0$ spectra. Our preliminary test therefore suggests that the model has some generalizability, at least in terms of such a cross-redshift application. 

\subsection{2D "Bubbles"}
We expect the regions surrounding the first bright sources of hard ionizing radiation (quasars) to be
spatially distinct (e.g., \citealt{mcquinn11,laplante17}). These "bubbles" will be initially heated to
higher temperatures than their surroundings, rising from $\sim 10^{4}$ K to $\sim2\times10^{4}$ K 
(e.g., \citealt{furlanetto08,puchwein2015}). In order to approximate this and show visually the
possibilities involved in temperature reconstruction, we have applied the increased temperatures used in 
previous tests, but this time expanding to multiple sightlines. We still use the NN to reconstruct the 
temperatures on an individual sightline basis, but have picked the background sources to lie in a line, so
that the sightlines sample a full plane through the simulation volume. Observationally, this arrangement of
sightlines would not be possible of course, but one could imagine using densely packed sightlines
such as those used in IGM tomography (e.g., \citealt{schmidt19,newman20,horowitz21}) to map
out the temperature.

\label{2Dbubble}
\begin{figure*}
    \begin{center}
    \begin{subfigure}[b]{\columnwidth}
        \includegraphics[width=0.9\textwidth]{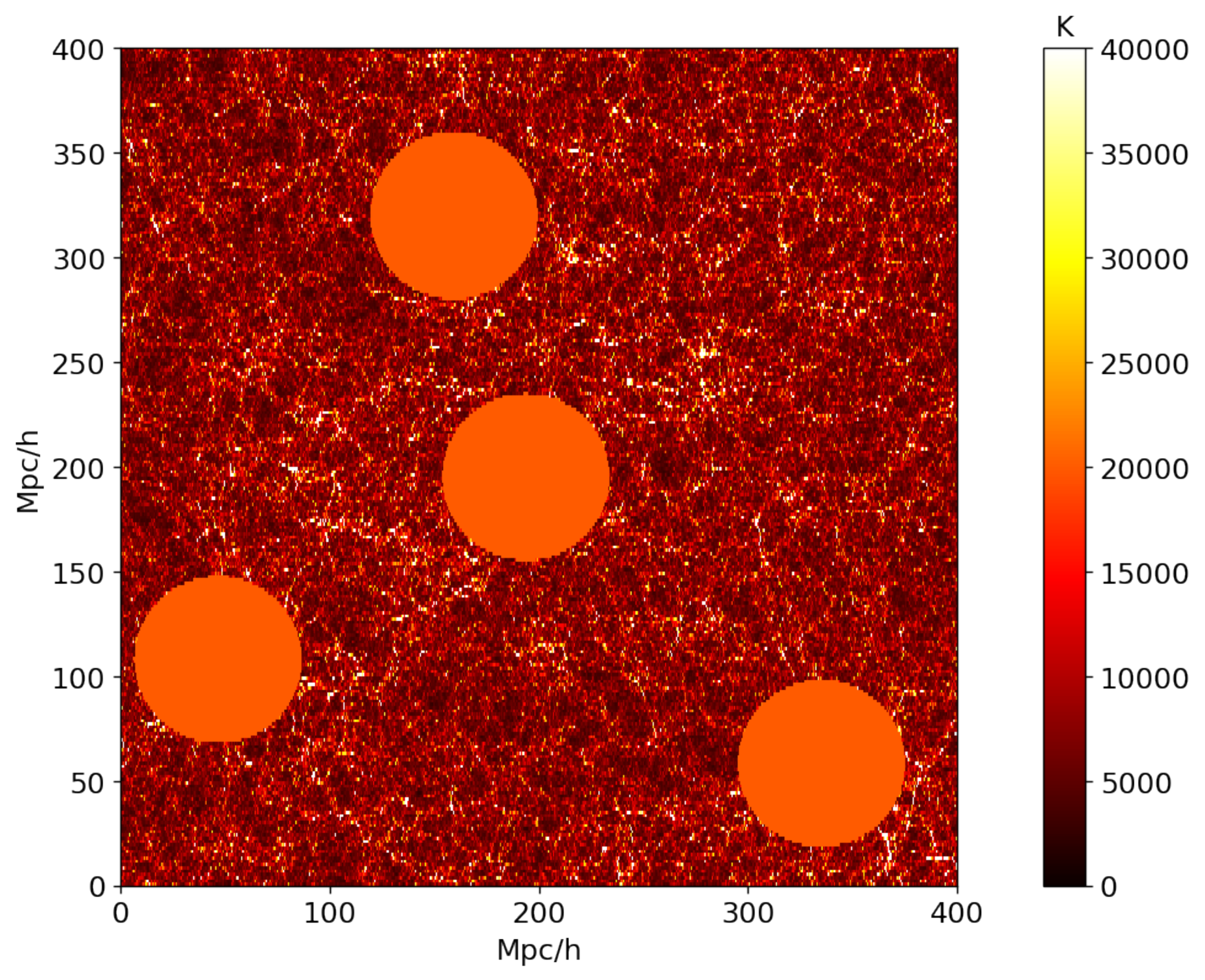}
    \end{subfigure}
    \begin{subfigure}[b]{\columnwidth}
       \includegraphics[width=0.9\textwidth]{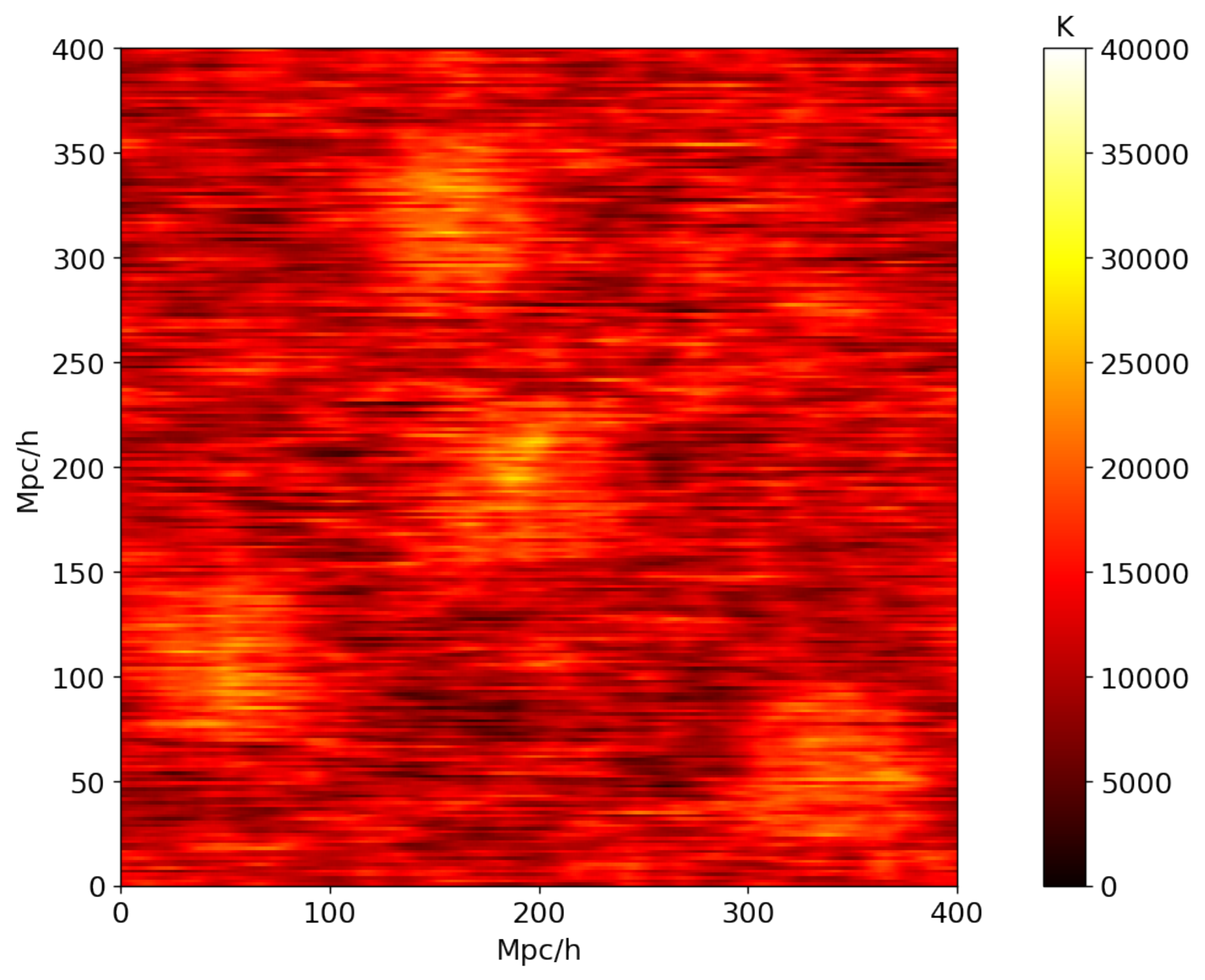}
    \end{subfigure}
    \begin{subfigure}[b]{\columnwidth}
       \includegraphics[width=0.9\textwidth]{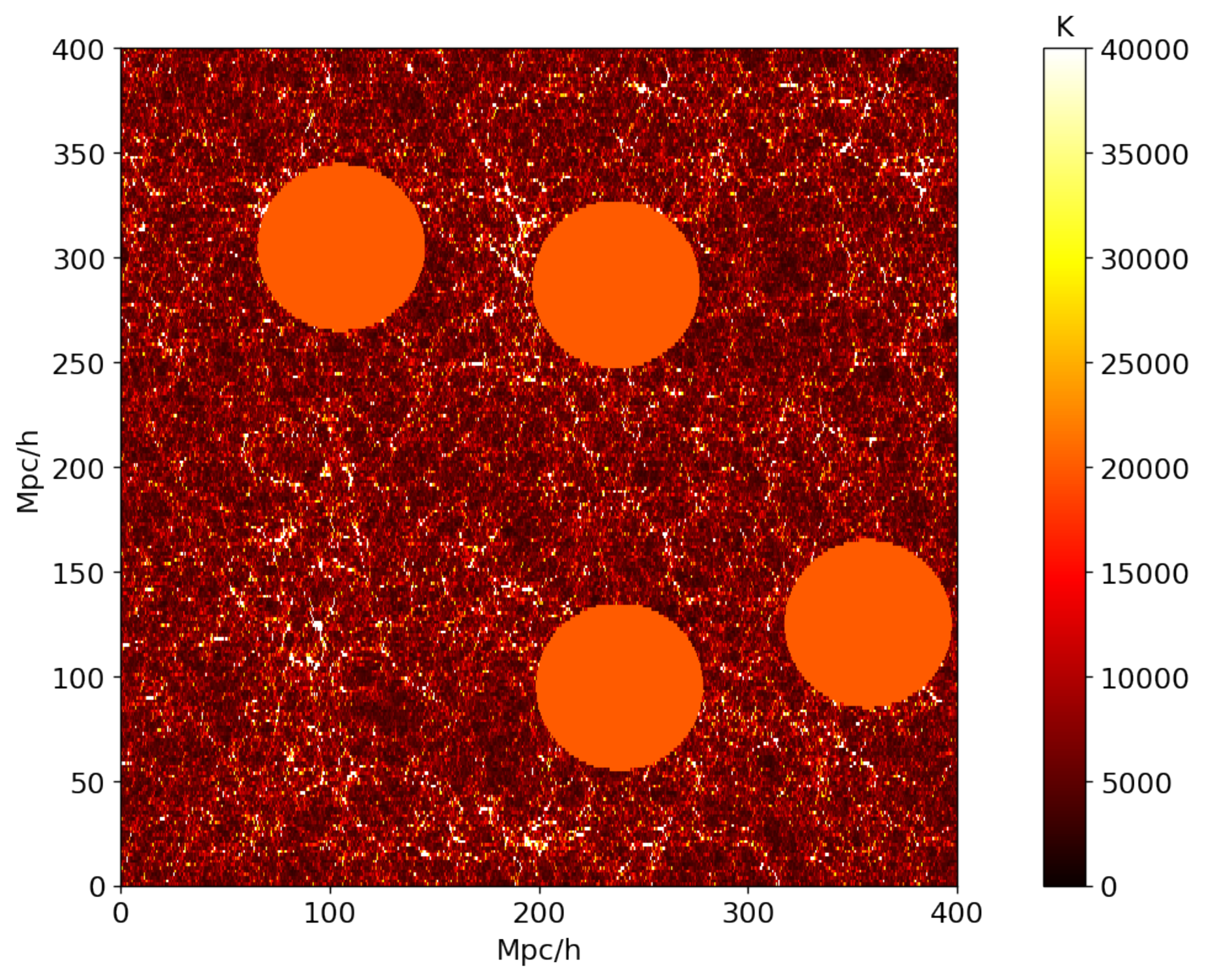}
    \end{subfigure}
    \begin{subfigure}[b]{\columnwidth}
       \includegraphics[width=0.9\textwidth]{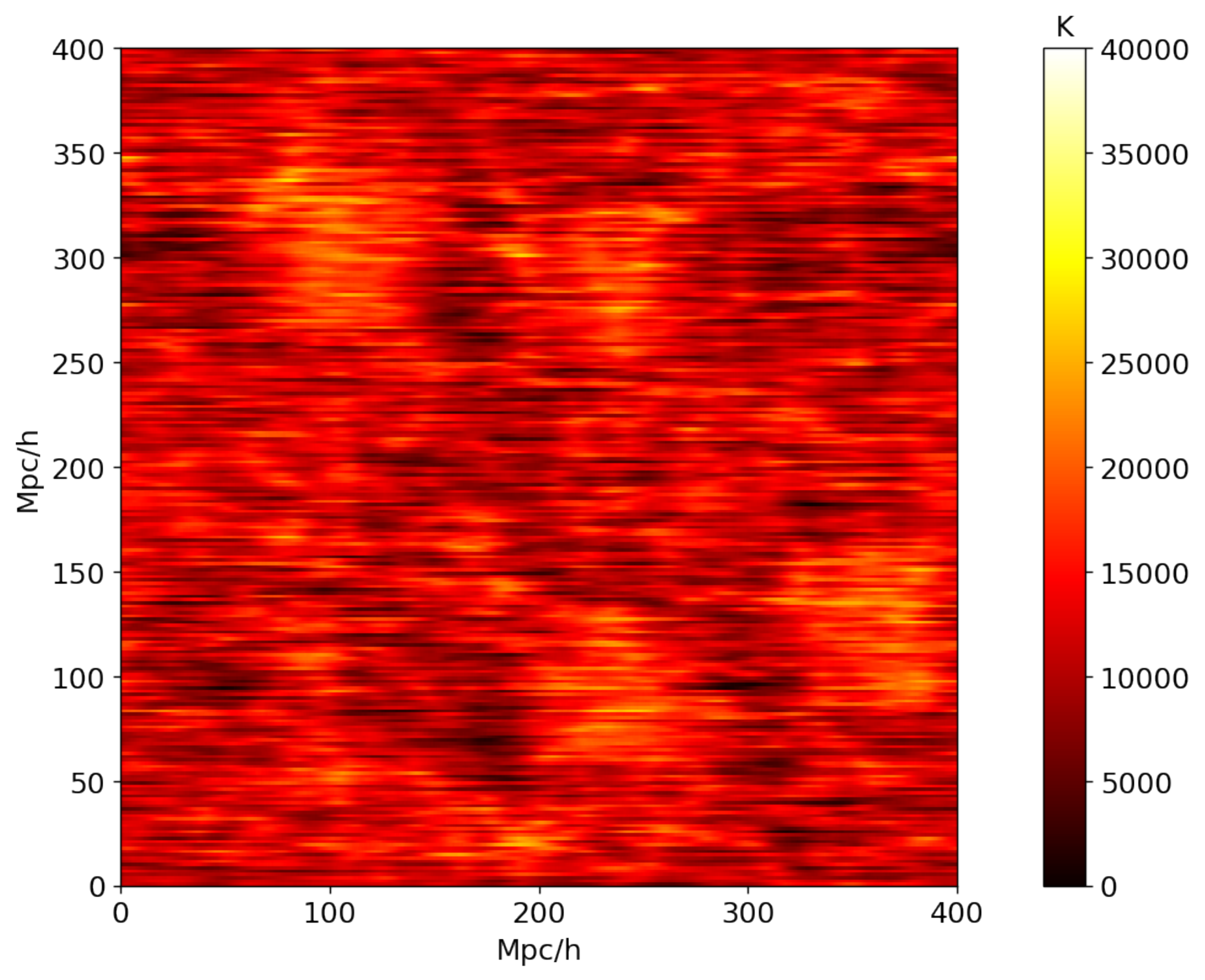}
    \end{subfigure}
    \caption{Actual versus NN reconstruction of temperature in a 2D plane
    through the simulation with mock quasar-dominated region with constant temperature $T_c = 20,000$ K and radius $r=40 \hmpc$. The locations of the "bubbles" are randomly generated. 
    The NN reconstruction was carried out on individual sightlines (running parallel to the $x$-axis) in the plane.
    The color scale saturates so that pixels with $T>40,000$ K are white. Top left: actual temperature map for $z=3.0$ Top right: reconstructed temperature map for $z=3.0$. Bottom left panel: actual temperature map for $z=2.5$. Bottom right panel: actual temperature map for $z=2.5$}
    \label{2DBubbles1}
    \end{center}
\end{figure*}
\begin{figure*}
    \begin{subfigure}[b]{\columnwidth}
        \includegraphics[width=0.9\textwidth]{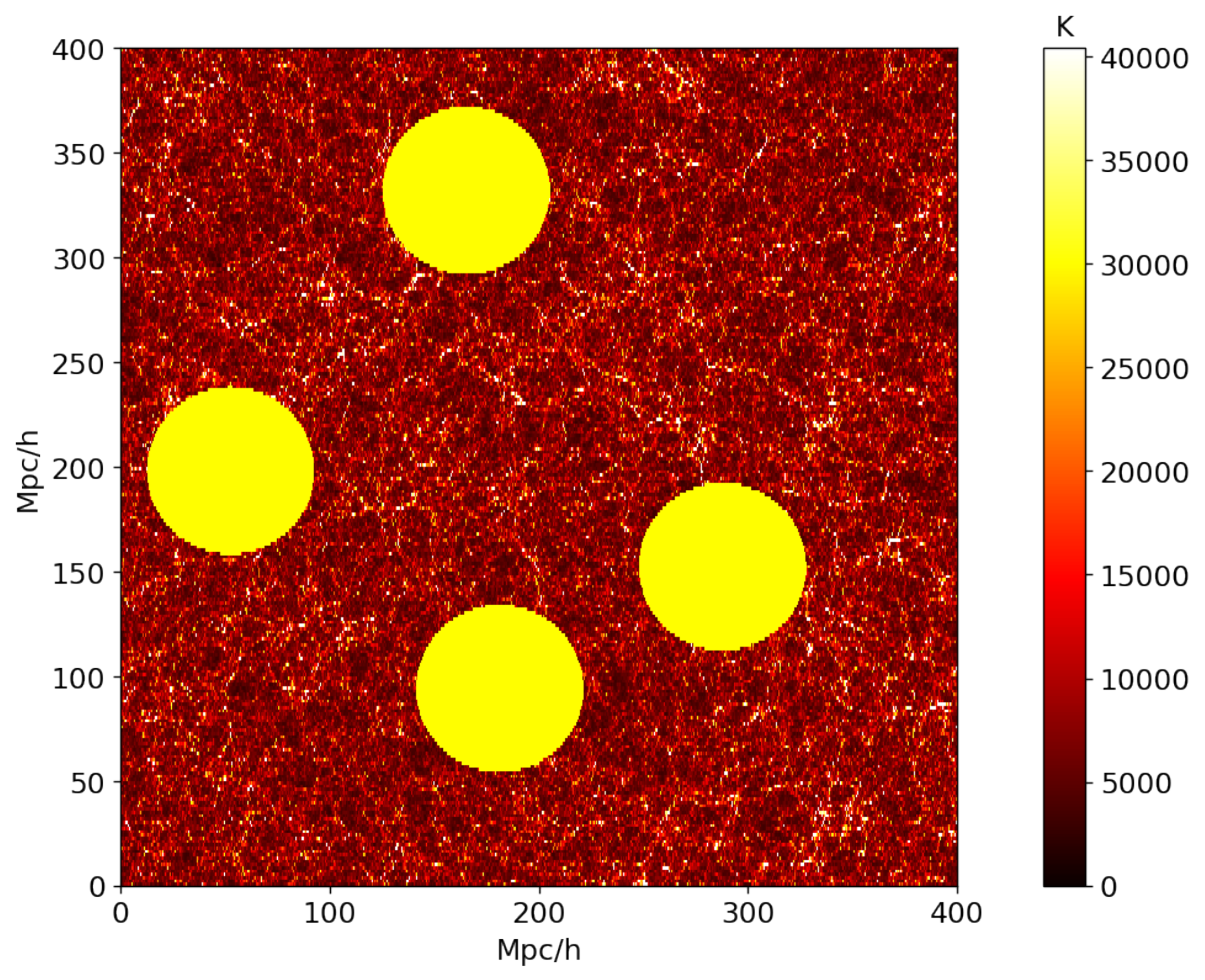}
    \end{subfigure}
    \begin{subfigure}[b]{\columnwidth}
       \includegraphics[width=0.9\textwidth]{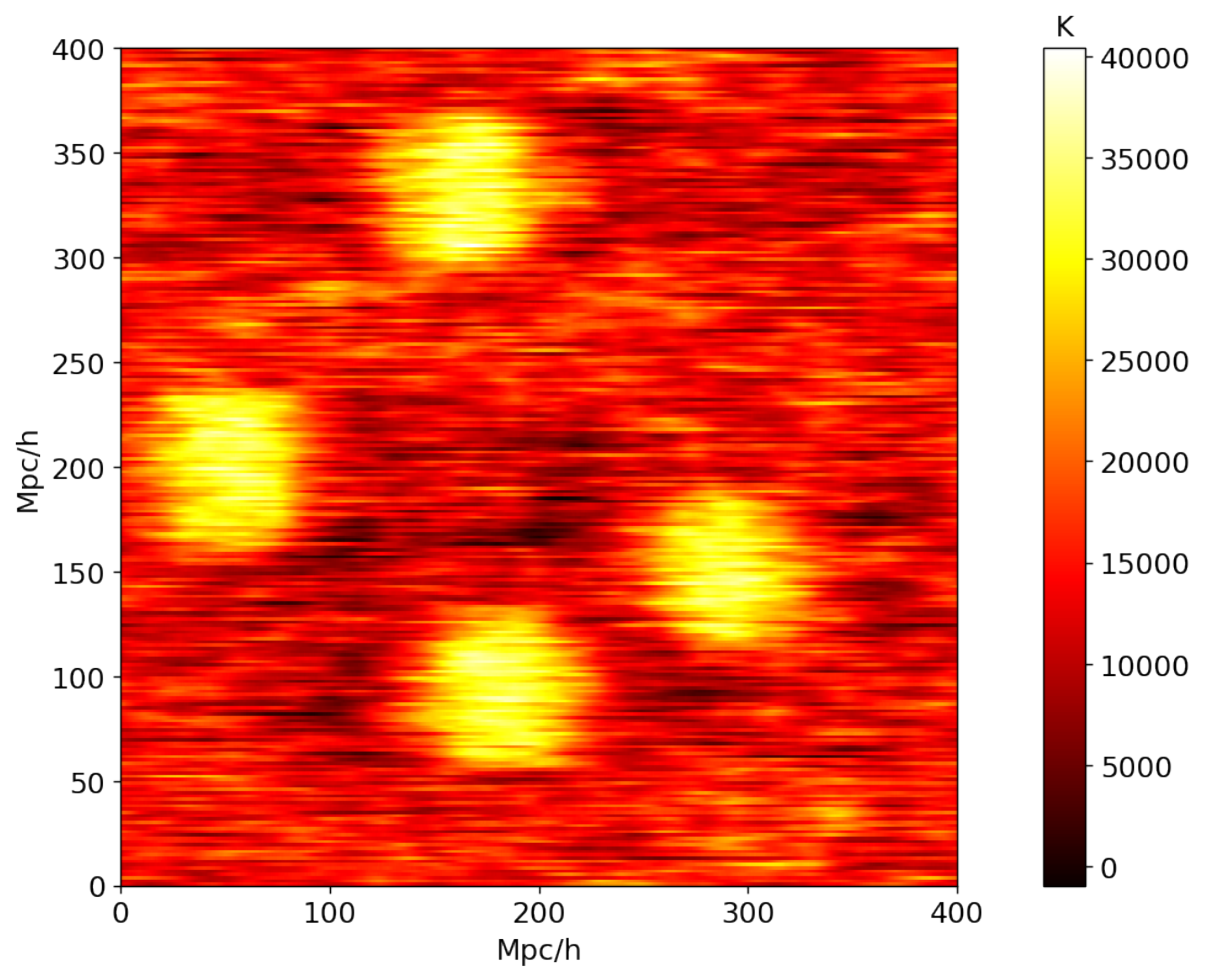}
    \end{subfigure}
    \begin{subfigure}[b]{\columnwidth}
       \includegraphics[width=0.9\textwidth]{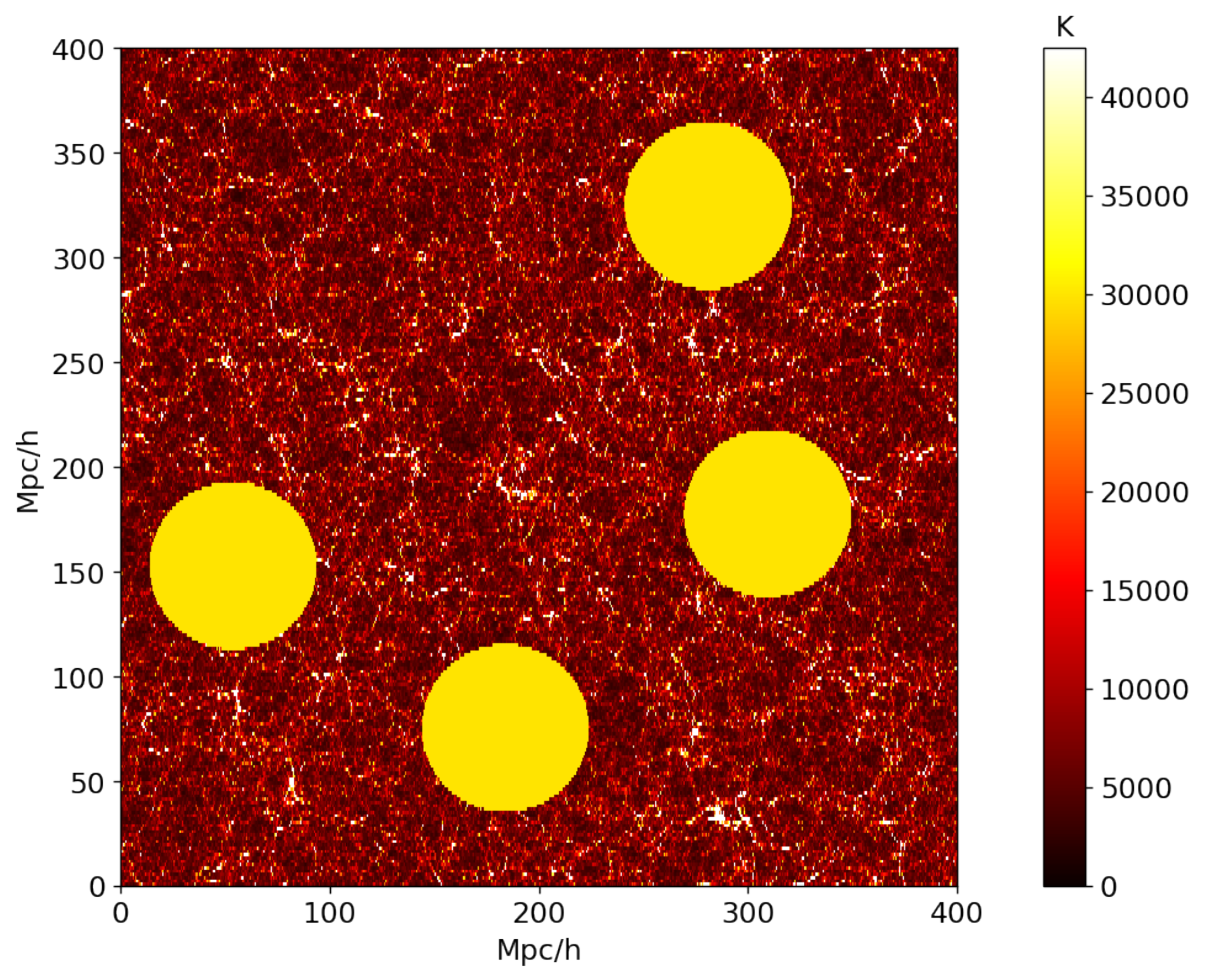}
    \end{subfigure}
    \begin{subfigure}[b]{\columnwidth}
       \includegraphics[width=0.9\textwidth]{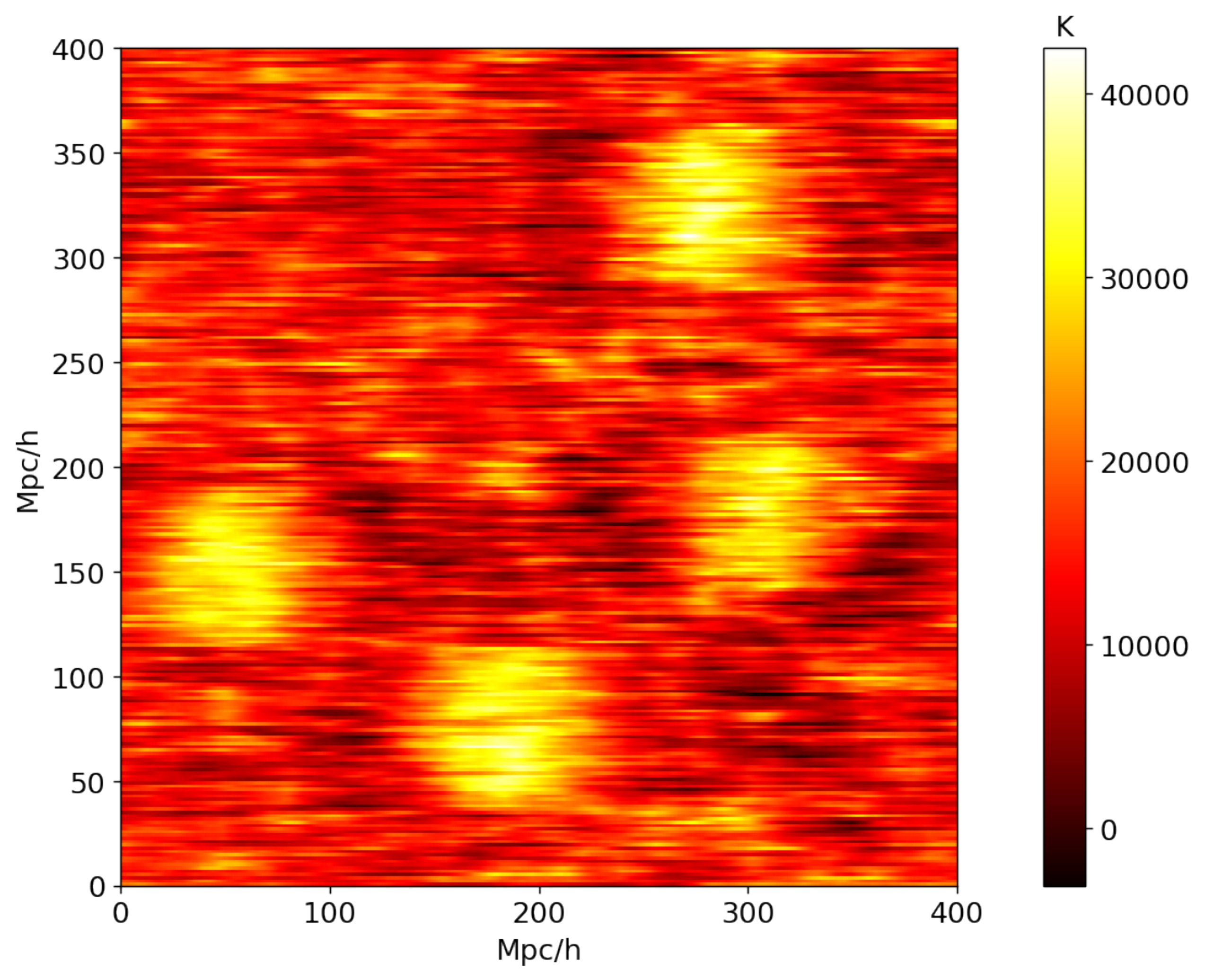}
    \end{subfigure}
    \caption{Actual versus NN reconstruction of temperature in a plane through the simulation with mock quasar-dominated regions with constant temperature $T_c = 30,000K$ and radius $r=40 \hmpc$. The locations of these "bubbles" are randomly generated. The NN reconstruction was carried out on individual sightlines (running parallel to the $x$-axis) in the plane. The color scale saturates pixels with $T>40,000K$ to be white. Top left: actual temperature map for $z=3.0$. Top right: reconstructed temperature map for $z-3.0$. Bottom left panel: actual temperature map for $z=2.5$. Bottom right panel: actual temperature map for $z=2.5$}
    \label{2DBubbles2}
\end{figure*}

 We randomly select this slice of 256 sightlines lying in a plane in the original simulation cube.
 We show the geometry in Figure \ref{2DBubbles1}, where the $x$-axis is distance along the line of sight (distance from the observer), and the $y$-axis is the width spanned by the 256 sightlines. The square has the same dimensions of the cube with side length $400 \hmpc$. We then impose four quasar-dominated regions with constant $T_c=20,000$ K or $T_c=30,000$ K for both redshifts $z=2.5$ and $z=3.0$ (i.e,
 four different cases). 

Figure \ref{2DBubbles1} and \ref{2DBubbles2}  show the actual temperature (top left and bottom left panels) and the Neural Network reconstructed temperature (top right and bottom right panels), color-coded based on the temperature value. For both figures the locations of the "bubbles" are randomly generated, and the radius of the bubbles are $40\hmpc$, $1/10$ of the box length $l=400 \hmpc$. We apply the models that are trained to predict the temperature along individual sightlines with quasar-dominated patches with $67 \hmpc$ and $T_c = 20,000$ K and $T_c = 30,000$ K, respectively. The NN was therefore not trained specifically to look for heated regions in spectra of the same size that occur in the bubbles.

The NN reconstructions for the individual sightlines are then put together to form the 2D  reconstruction, also color-coded by temperature value (the right column of figures). For Figure \ref{2DBubbles1}, $T_c=20,000$ K, and for Figure \ref{2DBubbles2} we show the cases where $T_c = 30,000$ K . For both figures, the spectra in the top row have $z=3.0$, and the second row $z=2.5$. The color scales for the 8 panels are the same, with the white pixels where the temperature is above $40,000$ K being saturated.

Qualitatively, we observe that for both redshifts the locations of the bubbles are correctly placed in the prediction, visually distinct from the background. The shapes of the "bubbles" in the predictions are very roughly circular, but on the edge of each ``bubble", the width of the distinct high temperature regions in the reconstruction has larger error. As expected, the predicted "bubbles" are more distinct from the background for the higher $T_c = 30,000$ K. That the prediction is less accurate for sharper and lower mock quasar regions is consistent with the trend in the reconstructed energy ratio discussed in Section \ref{energyRatio}.

Since the loss function measures the deviation in results for the whole sightline, we expect in the 2-D reconstruction some recovery of temperature structures outside the bubbles that resemble
temperatures in the rest of the IGM. Indeed there are qualitatively a few places where thin long structures in the background are reconstructed, for example of the contrast between a higher temperature region and a lower one in the lower left portion in the bottom two panels in Figure \ref{2DBubbles2} ($z=2.5$ and $T_c=20,000$ K), and the middle portion the bottom two panels in Figure \ref{2DBubbles1} ($z=2.5$ and $T_c=30,000$ K). Since in the training the temperature spectrum is first smoothed by a Gaussian function with kernel size of 48 pixels ($4.69 \hmpc$, as mentioned in section \ref{sim}), the 2D reconstructions are smoother than the true temperatures.

\section{Discussion and conclusions}
\label{DiscusssionandConclusions}
\subsection{Conclusions}
We have trained a convolutional neural network to recover the smoothed temperature distribution from 
simulated \lya forest spectra. In our tests of this, we have concentrated on the ability to find
distinct regions or patches of high temperature that could be caused by helium reionization heating
around bright sources. Our conclusions are the following:
\begin {enumerate}
\item{NN are able to decode some
 information about the IGM
temperature from the structure of the \lya\ forest, using Doppler broadening information.} 
\item{For distinct heated regions, the performance depends on the widths and the temperature of the region. An increase in widths and/or temperature generally implies a higher reconstruction accuracy.}
\item{The model can be applied to spectra with Gaussian noise added (minimum S/N=20 tested) with only a slight decrease in accuracy.}
\item{The reconstruction applied to a different redshift than was used for NN training
also does not lose much predictive power.}
 
\end{enumerate}

\subsection{Discussion}
\label{discussion}

\subsubsection{Potential improvements to NN architecture}

As seen in Figure \ref{archi}, our Neural Network architecture does not have much complexity, with only one convolutional layer and two fully-connected linear layers. Given that we have been experimenting on the same model, future work will involved attempting to use more sophisticated architectures. The challenge in increasing the number of layers is the risk of overfitting. As discussed in section \ref{NN}, one more fully-connected linear layer in the current structure would fail to converge. 

If we cannot overcome overfitting by changes in architecture alone, another possibility is to increase the size of the input data. If we take more than one spectrum as the input, for example capturing information about nearby pixels in three-dimensions, this would feed more 
information into the NN at once.
Recently, \cite{muller21} have shown how the use of such tomographic data can lead to improvements
in temperature measurement from the forest. Because the source positions are unstructured (sightlines
do not lie on a regular grid, but are essentially randomly distributed), this will require development of analysis techniques
that that take this into account (e.g., \citealt{metcalf20}).

\subsubsection{Potential improvements to NN training}
\label{discussionTraining}
As mentioned in Section \ref{preprocess}, we have briefly attempted different training procedures,
by 
artificially imposing a random $\rho-T$ relation on each spectrum in the training set. This is
instead of our fiducial training, where the spectra follow the $\rho-T$ relation from the hydrodynamic simulation except for randomly placed hot bubbles.
The idea behind trying a different training is for the model to better recognise Doppler broadening effects in a more general set of situations. We set $T= \textrm{B} \rho^k + \textrm{fluctuations}$, where for each individual sightline, we randomly select B (temperature at mean density) from $(3000\textrm{ K}, 25000\textrm{ K})$ and set $k=1-\tanh(\textrm{B}/10000)$. We also apply mock quasar-dominated regions. After this training the NN is able to reconstruct the temperature
for any spectrum in a test set with a randomly generated $\rho-T$ relation, with 
a mean fractional rms error  per pixel of $0.19 \pm 0.13$.  This example is merely illustratory, but shows how the
NN techniques could be extended beyond looking for hot bubbles in
the IGM temperature distribution.

\subsubsection{General discussion}

We have shown that NN are able to infer temperature information
from the broadening of \lya\ forest absorption features. This is merely
a preliminary investigation, and there are several caveats.
As with all NN studies, we are restricted by the training set employed.
We have relied on a simulated \lya\ forest training set from 
a hydrodynamic cosmological simulation with a single set of CDM cosmological
parameters. We expect that different cosmologies, for
example, fuzzy dark matter (FDM, \citealt{hu2000}) would have structures
which are intrinsically smoother, so that the NN trained on a CDM simulation
could misinterpret the spectral information and yield biased values
of the temperature. Indeed, the \lya\ forest has been used to constrain the
mass of FDM particles \citep{irsic17}, where the effect of
IGM temperature needs to be accounted for.
In the case of NN constraints, one should increase the training set to include other
simulations, with different parameters.

We have tried a limited version
of this, by using a set of spectra from  a different redshift to train
the NN. This results in slightly less accurate measures of the absolute temperature,
but temperature differences in different regions, such as might occur in
hot bubbles, are still easily detectable. In the future, one will need to
ensure that the underlying model used to train the NN has more flexibility,
including different cosmological parameters, and these could be
jointly constrained with the temperature (see e.g., \citealt{villa21}
for recent efforts to measure cosmological and galaxy formation
physics parameters from simulation outputs using DL).

The IGM temperature structure models we have tested are toy models, spherical bubbles with an isothermal equation of state. In practice,
the situation in the Universe is expected to be different and more
complex (\citealt{upton20,villasenor21}).
High density regions surrounding quasars will undergo HeII reionization
earlier than lower density parts of space (through the action of photons
with a short mean free path). They will then cool as the Universe
moves to lower redshift, and because of this will be cooler than their
surroundings which reionize later. Depending on the redshift of observation,
cooler bubbles might be a more accurate structure to search for. This
is unlikely to prove a problem for the NN method, as we have shown
that low temperature regions (e.g., the unheated parts of spectra
shown in Figures \ref{exFT} and \ref{exampleConstQuasar}) can also have their thermal properties
constrained.

An additional issue is that the effects of extra heating in our study were only added in post processing to spectra, as thermal broadening. In the real Universe, photo-heating will change the gas pressure, and self consistent modeling would include hydrodynamics (e.g., as carried out by \citealt{laplante17}). Including information on adjacent sightlines could in principle allow the NN to disentangle the effects of pressure smoothing which occur in three dimensions, and thermal broadening along the line-of-sight (see e.g., \citealt{peeples10,muller21}).

Although capturing thermal broadening features for this type of work requires high resolution,
echelle spectra of the \lya\ forest (e.g., \citealt{gaikwad21}), we have shown  that both signal to noise
ratios of 20 and 50 allow high temperature regions to be found and characterised. The
combination of Deep Learning, simulations and observations 
may
therefore become a useful tool in our quest to understand the thermal history of the IGM.

\subsection*{Acknowledgements}
We thank Lawrence Huang and Hitesh Arora for useful discussions and for making their \lya\ reconstruction
code publicly available.
RW acknowledges support from the NSF AI Institute: Physics of the Future Summer Undergraduate Research Program in Artificial Intelligence and Physics.
RACC is supported by  NASA ATP 80NSSC18K101,  NASA ATP NNX17AK56G, NSF NSF AST-1909193, and the NSF AI Institute: Physics of the Future, NSF PHY- 2020295.

\subsection{Data availability}
The simulated spectrum data underlying this article is accessible through a code repository.\footnote{\tt https://github.com/lhuangCMU/\linebreak deep-learning-intergalactic-medium. } or through request 
to the authors.

\bibliographystyle{mnras}
\bibliography{ref} 

\end{document}